\documentclass{biophys}
\usepackage{helvet,times}
\usepackage{bm,textcomp}

\jno{kxl014} 
\gridframe{N}
\cropmark{N}

\doi{doi: }

\usepackage{amsmath}
\usepackage[round,numbers,sort&compress]{natbib}

\begin{document}

\setcounter{page}{1}

\title{Quantifying the role of chaperones in protein translocation by computational modelling}

\author{Salvatore Assenza, Paolo De Los Rios, and Alessandro Barducci}

\address{Laboratoire de Biophysique Statistique, Ecole Polytechnique F\'ed\'erale de Lausanne (EPFL), CH-1015 Lausanne, Switzerland}

\begin{abstract}
{The molecular chaperone Hsp70 plays a central role in the import of cytoplasmic proteins into organelles,
driving their translocation by binding them from the organellar interior. Starting from 
the experimentally-determined structure of the \textit{E. coli} Hsp70, we computed, by means of molecular simulations,
the effective free-energy profile for substrate translocation upon
chaperone binding. We then used the resulting free energy to quantitatively characterize the kinetics of the import 
process, whose comparison with unassisted translocation highlights 
the essential role played by Hsp70 in importing cytoplasmic proteins.}
{Insert Received for publication Date and in final form Date.}
{Address reprint requests and inquiries to S. Assenza (salvatore.assenza@epfl.ch)
or A. Barducci (alessandro.barducci@epfl.ch).}
\end{abstract}

\maketitle 
\section*{Introduction}
Molecular chaperones are protein machines that assist other proteins in 
various cellular processes.
70-kDa Heat Shock Proteins (Hsp70s) are possibly the most versatile chaperones,
supervising a wide variety of cellular tasks \cite{mayerbukau}
that range from disaggregation of stable protein aggregates \cite{aggr1} to 
driving posttranslational import of 
cytoplasmic proteins into organelles \cite{neupert, rapoportER, theg}.
Notably, Hsp70s play a fundamental role in the import of proteins into
mitochondria \cite{neupert}. Indeed, the majority of mitochondrial proteins are actually post-translationally imported 
from the cytosol through outer (TOM) and inner (TIM) membrane pore complexes \cite{neupert}. According to the current view,
an ATP-consuming import motor located into the mitochondrial matrix drives the inward translocation of nuclear-encoded proteins. 
Mitochondrial Hsp70 (mtHsp70) is the central element of this motor: it is recruited by TIM 
on the matrix side and it binds the incoming protein upon ATP hydrolysis, thus 
driving its translocation.

The structure of Hsp70 is highly conserved \cite{mayerreview} and consists of two 
large domains connected by a small flexible linker (see 
fig.1). Specifically, the Nucleotide Binding Domain (NBD)
is the ATPase unit of the chaperone, while the Substrate Binding Domain (SBD)
directly interacts with specific sites on the incoming protein. These binding sites are frequently found 
in protein sequences, so that multiple chaperones are likely to bind the same substrate.

The precise mechanism by which Hsp70 exerts its pulling action
has been debated in the literature and several models have been proposed \cite{neupert,powerstroke,delos}.
The \textit{Brownian ratchet} \cite{neupert} assumes that, thanks to the chaperone large size, Hsp70 binding prevents 
the retrotranslocation of the substrate into the pore, thus biasing the random fluctuations toward the matrix.
Alternatively, according to the \textit{power stroke} \cite{powerstroke} the chaperone actively pulls the incoming
protein by using TIM as a fulcrum. 
Later, according to the \textit{entropic pulling} model \cite{delos}, it was shown that 
an active force naturally emerges from a realistic physical description 
of the Brownian ratchet, 
thus reconciling the two views \cite{paoloreview}. Indeed, 
the excluded volume of the chaperone, besides preventing
retrotranslocation, reduces the conformational space available to the incoming protein,
thus decreasing its entropy. 
This reduction depends on the 
length of the imported fragment of the substrate, therefore resulting in a free-energy 
gradient which favors the import. 

In the present work, we evaluate this thermodynamic force in an effective one-dimensional space 
where the state of the system is represented by the number $n$ of imported residues. 
In order to do so, for each value of $n$ we compute the
effect of chaperone binding on the free energy of the system
by means of coarse-grained Molecular Dynamics (MD) simulations. 
This result is then used 
to devise a simplified yet quantitative analysis of the import, 
described as a one-dimensional diffusion process on the computed free-energy landscape.

\begin{figure}[t!]\vspace*{3pt}
\centering{\includegraphics[scale=0.25]{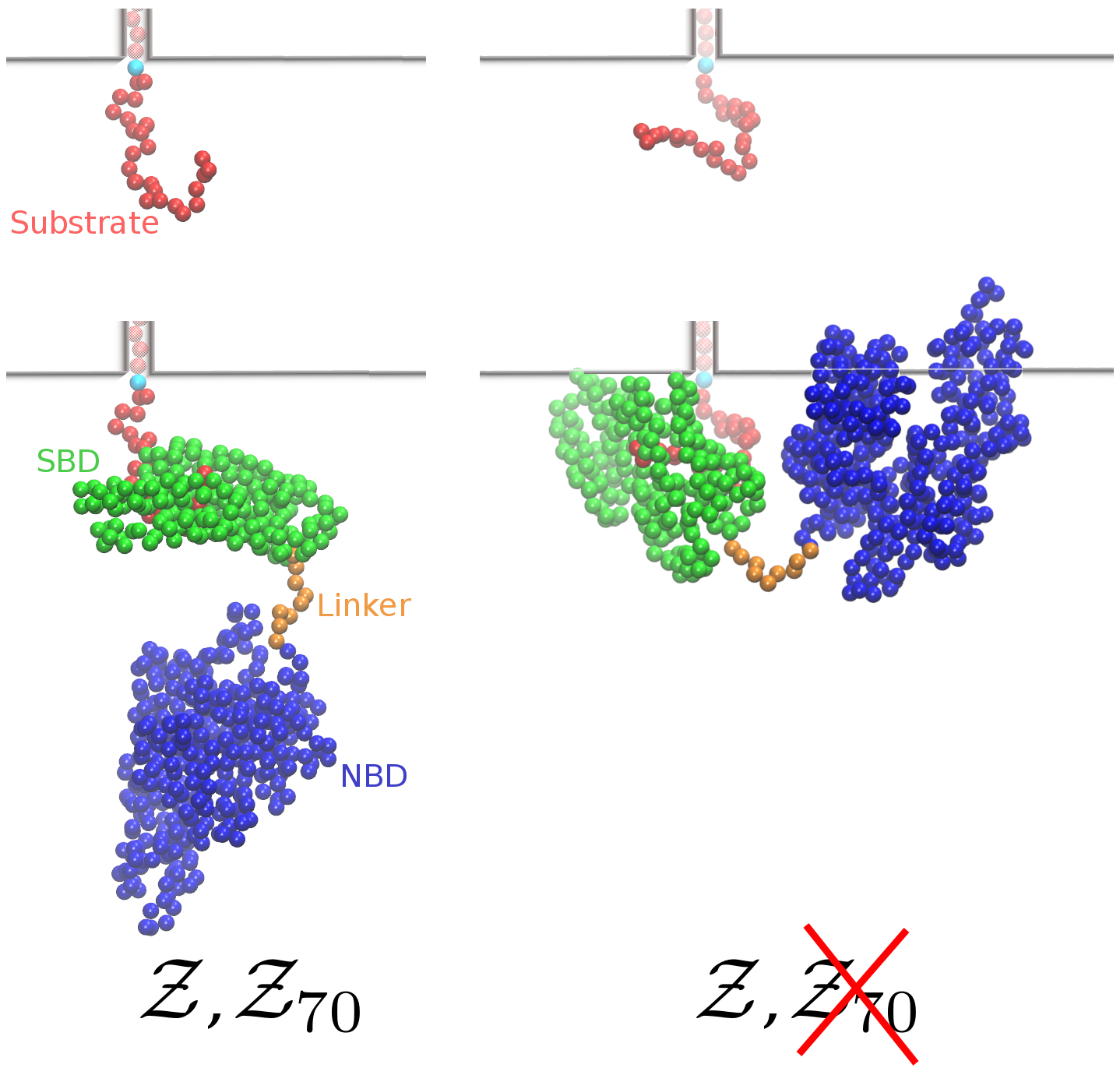}}\vspace*{-5.5pt}
\caption{
Two representative conformations of the substrate without (top) and with (bottom) bound chaperone.
While in the absence of Hsp70s both conformations contribute to $\mathcal{Z}(n)$, upon chaperone binding the one on the right is not taken into account
in $\mathcal{Z}_{70}(n)$ due to sterical clash with the wall.
The $n$th residue of the substrate,
which is constrained on the wall, is colored in cyan. The shaded beads
inside the channel are here drawn only for representative purposes.}\vspace*{-8pt}
\end{figure}
\section*{Methods}
\subsection*{Details of MD simulations}
We coarse-grained both the substrate and the chaperone 
by considering one interaction site per residue centered on the C$_\alpha$
atom. Residue-residue excluded-volume interactions were 
modeled with a repulsive Lennard-Jones potential with parameters $\sigma=3.8$ \AA$\,$
and $\varepsilon=3k_BT$.
The substrate was modeled by using the local flexible potential introduced in \cite{forcefield}. In particular, we 
made use of the functions denoted as O-X-Y and X-X for the bending and torsional contributions, respectively.
The experimental structure of ADP-bound Hsp70 \cite{bertelsen} (PDB: 2KHO) was used to model the chaperone. 
In particular, the NBD (residue 4 to 387) and SBD (residue 397 to 603) 
were treated as rigid bodies, while the flexibility of the interdomain linker 
was accounted for by means of the potential described above.
In order to 
reproduce a correct chaperone-substrate arrangement, we took advantage 
of the substrate-bound X-ray structure of DnaK SBD (PDB: 1DKX \cite{nbd}).

MD simulations were performed using LAMMPS code \cite{lammps} at constant temperature ($T=300$ K) by means of a Langevin thermostat
with damping parameter equal to $100$ fs and using an integration timestep of $10$ fs.
For each value of $n$ in the range $8\leq n\leq26$ we performed a MD simulation of $5\cdot10^{10}$ timesteps (examples of the 
convergence of the ratio $\mathcal{Z}_{70}(n)/\mathcal{Z}(n)$ are reported in the Supplementary Material), and 
the error on the free-energy profile was estimated by block averaging \cite{frenkel}.

\subsection*{Details of the Stochastic Simulations}
The import process
was simulated by means of a Monte Carlo (MC) algorithm driven by 
the free-energy landscape $F_{\mbox{\small import}}$, as determined from the sum of the
chaperones pulling contribution computed by means of the MD simulations and the 
unfolding free energy $F_u$. The latter is modeled as 
a sigmoidal function
\begin{equation*}
 F_u(n_{\mbox{\small in}})=\frac{F_u^{\mbox{\small max}}}{1+\exp\left[5-10(n_{\mbox{\small in}}-10)/\delta n\right]}\,,
\end{equation*}
where $n_{\mbox{\small in}}$ is the total number of imported residues, and $F_u^{\mbox{\small max}}$ and $\delta n$ are tunable parameters.
For a system at
position $n_{\mbox{\small in}}$, a trial move was attempted to either $n_{\mbox{\small in}}+1$ or
$n_{\mbox{\small in}}-1$ with equal probability and accepted 
according to the Metropolis criterion based on the free energy
$F_{\mbox{\small import}}$. 
To capture the sequence heterogeneity of the proteome, for each choice of $F_u^{\mbox{\small max}}$ and $\delta n$
we generated 25 independent binding site distributions, with the sole prescription 
that the average distance between consecutive binding sites was 35 residues as indicated by experiments \cite{rudiger}.
For every distribution we performed 10 independent realizations of the import 
process. 
Average import times were estimated from MC simulations, that correspond to 
overdamped Langevin dynamics when only local moves are considered \cite{vankampen,tianasutto}.
Rescaling the obtained import times by the acceptance rate, as proposed in \cite{marenduzzo}, did not affect the results, because 
of the large fraction of accepted moves observed in all the simulations ($>95\%$).

\section*{Results}
Protein import into organelles has been previously modeled as a one-dimensional stochastic 
process in the space of the imported residues \cite{rapoportmodello, elston1, elston2}.
In the present context, this protocol is justified by the timescale separation among 
substrate conformational dynamics, chaperone binding/unbinding and overall import. 
Indeed, the typical reconfiguration time of an unfolded protein ($\sim 100$ ns \cite{schuler}) is extremely fast compared to the experimentally-determined 
timescale for protein import into mitochondria (order of minutes \cite{lim}). Effects arising from substrate conformational dynamics, 
such as the chaperone-induced entropy
reduction, can be thus conveniently represented as effective free-energy profiles influencing the import dynamics. 
Moreover, the import timescale is also significantly slower than chaperone binding but faster than chaperone unbinding at physiological 
conditions. Indeed, according to the current understanding of the biochemical cycle of Hsp70 \cite{elife,mayerreview},
ATP-bound chaperones associate with the 
substrate with a timescale equal to $\sim 10^{-2}$ s 
(as estimated from a Hsp70-peptide association rate 
equal to $4.5\times 10^5$ $\mbox{M}^{-1}\mbox{s}^{-1}$ \cite{bindrate} and a chaperone concentration of $70$ $\mu$M in mitochondria \cite{chapconc}),
while dissociation takes place from the ADP-bound state, over timescales $\sim 10^3$ s \cite{mayerdiss}.
This suggests that, to our purposes, we can assume that a chaperone immediately 
and irreversibly binds each exposed binding site as soon as it is imported.
As a consequence, 
for the present purposes the number $n$ of substrate residues that have been imported into the mitochondrial matrix
is a convenient coordinate to describe the system, whose dynamics can be modeled as a diffusion process on the corresponding 
free-energy landscape.

\subsection*{Free energy calculation.}
The effect of the size of the chaperone is two-fold. On the one hand, bound Hsp70 prevents the retrotranslocation 
of the substrate beyond its binding point (Brownian ratchet model \cite{neupert}). On the other hand, the size of the chaperone leads also to a reduced number of sampled 
conformations (entropic pulling \cite{delos}), an effect not accounted for by the Brownian ratchet as it was originally conceived, but nonetheless 
intimately related to the same physical mechanism. For example, in the absence of Hsp70 the two substrate conformations shown in the top panel of 
fig.1 are both sterically allowed. However, upon chaperone binding the conformation on the right would result into
an overlap between the membrane and Hsp70 (bottom panel in fig.1), and it is therefore never sampled by the substrate when the chaperone is present. 
The free energy difference due to the loss of entropy is given by 
$\Delta F_c(n)=-k_BT\log\left(\mathcal{Z}_{70}(n)/\mathcal{Z}(n)\right)$, where 
$\mathcal{Z}_{70}(n)$ and $\mathcal{Z}(n)$ are the partition functions of the substrate with and 
without a bound chaperone, $k_B$ is the Boltzmann 
constant and $T$ the temperature (when enthalpic contributions are not taken into account, the partition functions
reduce to the number of sampled conformations, thus falling back to the original formulation 
of the entropic-pulling free energy \cite{delos}).
Here, we computed the free energy difference $\Delta F_c(n)$ by estimating the ratio $\mathcal{Z}_{70}(n)/\mathcal{Z}(n)$ 
for $n$ in $8\leq n\leq 26$ with 
multiple coarse-grained MD simulations. 
The substrate was modeled as a $n$-residues flexible chain with the position of the $n$th 
residue constrained on the inner mithocondrial membrane, 
represented here as a flat wall acting only on the substrate residues (see Methods for additional details). As a consequence,
the system could sample
\begin{figure}[t!]\vspace*{3pt}
\centering{\includegraphics[scale=0.2]{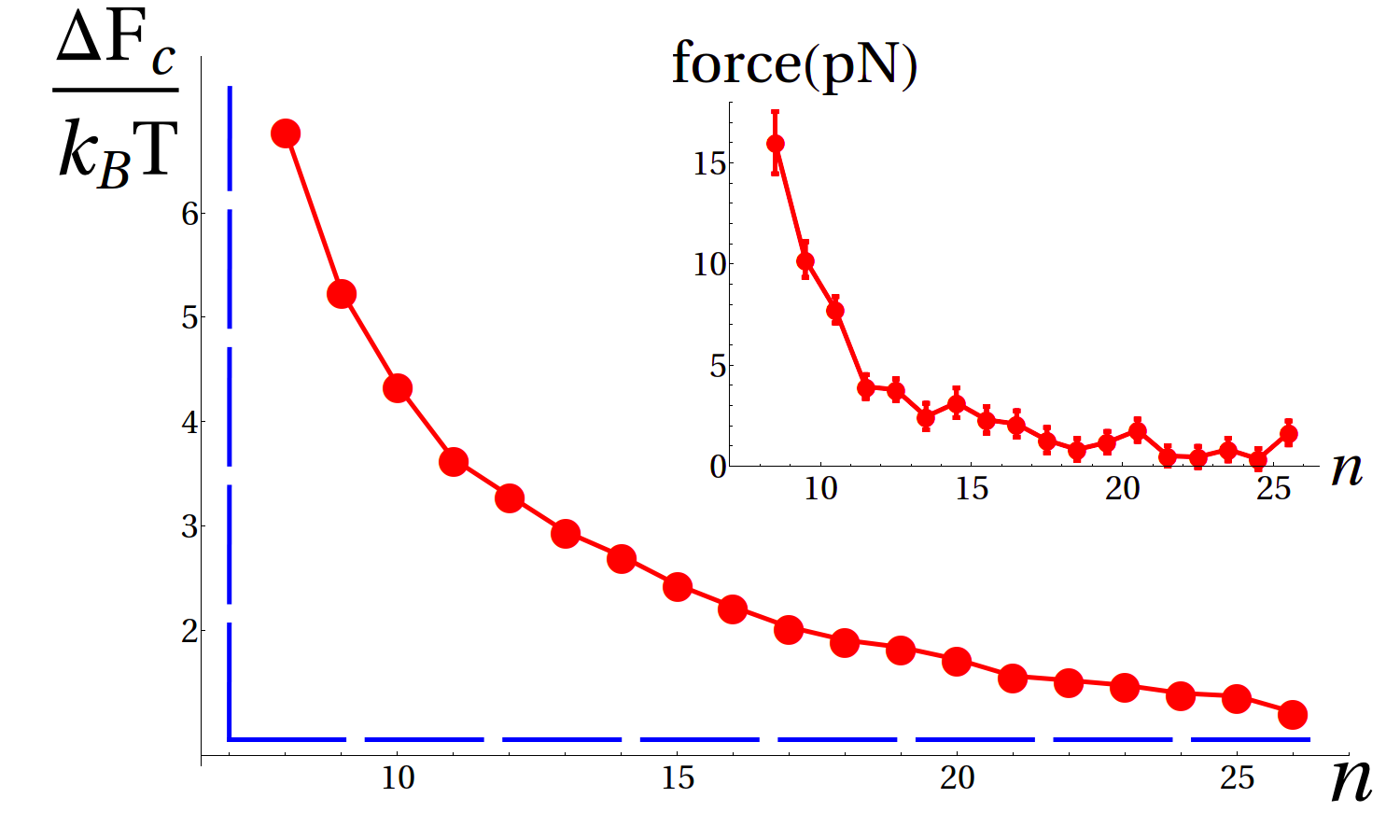}}\vspace*{-5.5pt}
\caption{
Free-energy profile due to chaperone binding as a function of $n$ (red circles). Error bars are smaller than 
the size of the symbols. The dashed blue line depicts the free-energy landscape predicted by the original Brownian ratchet, 
where the only effect of the chaperone is to prevent retrotranslocation beyond the binding site (infinite wall).
In the inset we report the thermodynamic force corresponding to the computed free-energy landscape.}\vspace*{-8pt}
\end{figure}
configurations involving an overlap between the membrane and the chaperone (see bottom-right panel in fig.1).
With this strategy, we could estimate the ratio $\mathcal{Z}_{70}(n)/\mathcal{Z}(n)$ as the fraction of time spent by the system 
in physically-acceptable, \textit{i.e.} non-overlapping, configurations.
Particularly, we focused on $n\geq 8$ in order to allow the exposure of a complete binding site.
From the computed values of $\mathcal{Z}_{70}(n)/\mathcal{Z}(n)$, we could retrieve the free energy 
$\Delta F_c(n)$ as a function of $n$, as reported in fig.2.
As expected, shorter imported fragments resulted into a larger fraction of rejected conformations, \textit{i.e.}
larger values of $\Delta F_c$, thus leading to a free-energy gradient favoring the import of the 
protein.
The slope of the entropic-pulling free-energy profile corresponds to the thermodynamic pulling force
exerted by a bound chaperone along $n$ (fig.2 inset). This force is in the piconewton range,
starting from around 15 pN 
and decreasing as $n$ increases. 
Remarkably, these results agree qualitatively with previous estimates based 
on strongly simplified representations of the system \cite{delos}, thus 
suggesting that comparable thermodynamic forces could be obtained 
by the same entropic pulling mechanism for macromolecules of similar size.

\subsection*{Stochastic simulations of the import process.} 
We modeled the import of 
cytoplasmic proteins
as a one-dimensional stochastic process depending on the number $n_{\mbox{\small in}}$
of imported amino acids. The effective free-energy profile guiding the system evolution results 
from protein unfolding \cite{schatz} and active chaperone pulling \cite{lim}. 
Assuming a two-states folding behavior,
a convenient choice to model the unfolding contribution to the free-energy landscape
is a tunable sigmoidal function $F_u(n_{\mbox{\small in}})$ (see Methods),
depending on two parameters that measure the total unfolding free energy ($F_u^{\mbox{\small max}}$) and the cooperativity of the unfolding process 
($\delta n$), with smaller values of $\delta n$ corresponding to higher cooperativity (top panel in fig.3). 
By tuning
these parameters, the formula can account for the wide variety
of imported proteins \cite{wilcox}.
The pulling action of the chaperone was modeled taking advantage of the free-energy profile  determined from molecular simulations. Particularly, we assumed here that:
i) Hsp70s associate with each binding site
as soon as it emerges from the pore, since they
are targeted at the TIM pore exit by specific interactions \cite{neupert};
ii) we considered only the contribution arising from the Hsp70 closest to the pore,
taking into account the relatively fast decrease of the
slope of $\Delta F_c$ (see fig.2) and the average frequency of binding sites (one every 35 amino acids \cite{rudiger}). 
Therefore, we added to the unfolding free-energy $F_u(n_{\mbox{\small in}})$ the chaperone contribution  
$\Delta F_c(n_{\mbox{\small in}}-n_B)$, with $n_B$ corresponding to the position of the binding site 
closest to the pore, measured from the matrix terminus of the substrate.

As an example, in the bottom panel 
of fig.3 we illustrate the evolution of the free-energy landscape during the import process of
a protein with $F_u^{\mbox{\small max}}=5k_BT$, $\delta n=100$ and two binding sites 
at $n_B=0$ (\textit{i.e.} at the matrix terminus) and $n_B=28$.
At the beginning of the import process, no chaperone is bound to the substrate and the import free energy is
simply given by $F_{\mbox{\small import}}(n_{\mbox{\small in}})=F_u(n_{\mbox{\small in}})$ (red
dashed curve). As soon as the first binding site is imported, a chaperone molecule binds the substrate and
its contribution $\Delta F_c$ is added to $F_u(n_{\mbox{\small in}})$ starting from the binding site $n_B=0$: 
$F_{\mbox{\small import}}(n_{\mbox{\small in}})=F_u(n_{\mbox{\small in}})+\Delta F_c(n_{\mbox{\small in}})$ (purple continuous curve). 
Finally, after the second binding site ($n_B=28$) is imported, another chaperone binds and 
the resulting free energy is
$F_{\mbox{\small import}}(n_{\mbox{\small in}})=F_u(n_{\mbox{\small in}})+\Delta F_c(n_{\mbox{\small in}}-28)$
(orange dot-dashed curve).
\begin{figure}[t!]\vspace*{3pt}
\centering{\includegraphics[scale=0.253]{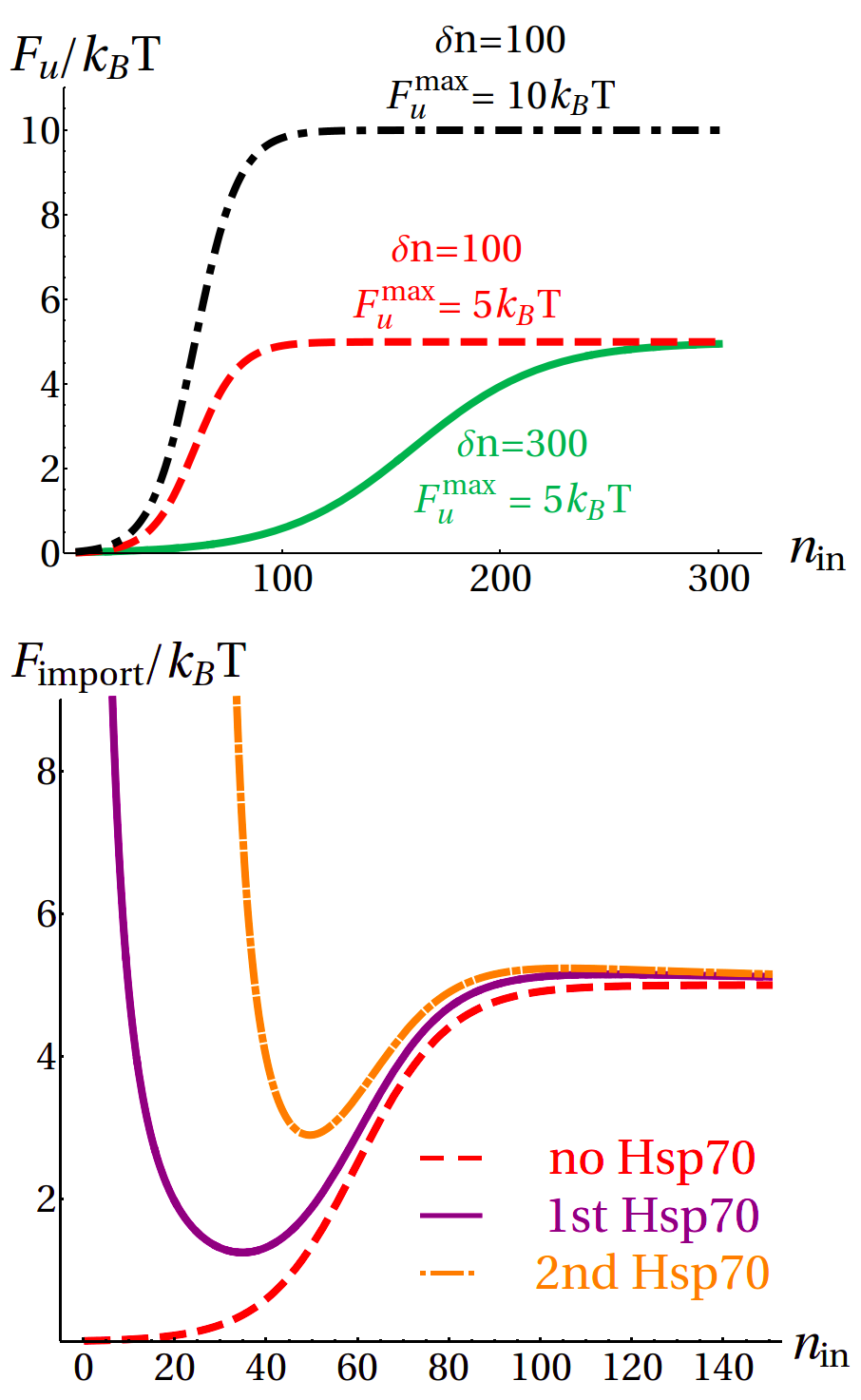}}\vspace*{-5.5pt}
\caption{Top: influence of the parameters $F_u^{\mbox{\small max}}$ 
and $\delta n$ on the unfolding free-energy.
Bottom: 
evolution of the total free-energy $F_{\mbox{\small import}}$ in a representative import process.}
\end{figure}
\begin{figure}[t!]\vspace*{3pt}
\centering{\includegraphics[scale=0.23]{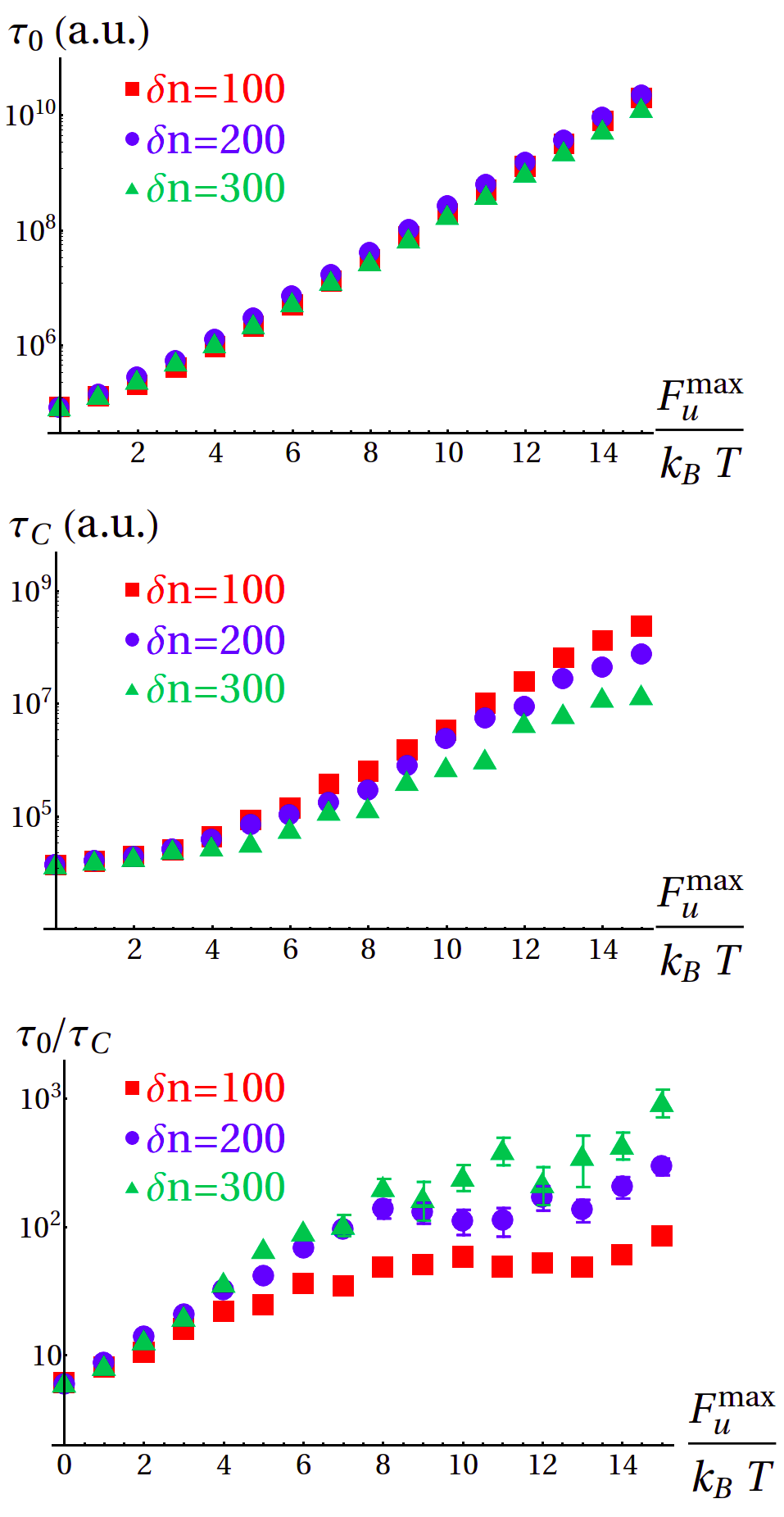}}\vspace*{-5.5pt}
\caption{Top: Average import times in the absence of chaperone ($\tau_0$) as 
a function of $F_u^{\mbox{\small max}}$ for different cooperativities (values for $F_u^{\mbox{\small max}}\geq 12k_BT$ were extrapolated 
by fitting the data in the range $4k_BT\leq F_u^{\mbox{\small max}}\leq 11k_BT$ with exponential functions). Center:
Average import times in the presence of Hsp70 ($\tau_C$) for the same cases as in the top panel.
Bottom: Acceleration of the process due to the assistance of Hsp70,
expressed as the ratio $\tau_0/\tau_C$.}\vspace*{-8pt}
\end{figure}

Following this approach, we computed the average import time (see Methods) of 300-residue proteins 
for  different values of $\delta n$ and a range of $F_u^{\mbox{\small max}}$
corresponding to the stability of a large fraction of the proteome \cite{dill}. 
In absence of Hsp70 assistance, the system must invariably overcome a free-energy barrier, and
the average import time $\tau_0$ 
increases exponentially with $F_u^{\mbox{\small max}}$, independently of cooperativity (fig.4 top).
In all the considered cases, the average import time for the chaperone-assisted process, $\tau_C$, is
sensibly smaller than $\tau_0$ (fig.4 center).
The chaperone pulling force reduces but does not completely eliminate the unfolding 
free-energy difference for stable proteins (large $F_u^{\mbox{\small max}}$), 
as in the case of the representative process shown in the bottom panel of fig.3.
In this regime, the import is still an activated process, and the average times increase exponentially 
with $F_u^{\mbox{\small max}}$.
Conversely, the pulling action of Hsp70 dominates over the unfolding contribution for marginally stable proteins
(small $F_u^{\mbox{\small max}}$), thus resulting 
in values of $\tau_C$ comparable to what found for 
the extreme case $F_u^{\mbox{\small max}}=0$.
The import kinetics is further modulated by $\delta n$, with
high cooperativity (small $\delta n$) resulting in longer translocation times.

In the bottom panel in fig.4 we illustrate the chaperone-induced kinetic advantage by reporting the ratio $\tau_0/\tau_C$.
This ratio ranges from a 10-fold gain for marginally stable proteins
to $10^3$ for extremely stable and noncooperative substrates, with 
the majority of the proteome ($F_u^{\mbox{\small max}}\geq 8k_BT$ \cite{dill}) accelerated at least 100 times.
If we take into account that protein import into mithocondria 
has been measured to happen in the timescale of several minutes \cite{lim}, our 
model indicates that the translocation process in the absence of chaperones would probably extend to hours or days.
Since such a slow process
would clearly be incompatible with the average lifespan of proteins 
and the duration  of the cell cycle, 
our results provide a molecular basis to support the essential role of chaperones in the 
\textit{in vivo} import process. 

\section*{Conclusions}
To summarize, in this work we derived a free-energy profile for the import process 
based on a molecular description of Hsp70
that rationalizes the requirement for chaperone assistance in mitochondrial protein import
observed in experiments.
The present results can be applied to other cases of Hsp70-driven translocation,
namely protein import into ER \cite{rapoportER} and chloroplasts \cite{theg}.
Moreover, this approach based on the combination of molecular simulations and 
kinetic modeling can be easily extended
to other Hsp70-mediated cell processes. In particular, this free-energy picture 
could help to understand some recent results pointing towards 
a fundamental role of Hsp70 in preventing the stalling of translation at ribosomes \cite{liuyan, lindquist}.
Owing to the universality of the interaction responsible for the effects studied here, namely excluded volume,
the same principles could apply to similar processes driven
by other biomolecules. 
\section*{AUTHOR CONTRIBUTIONS}
S.A., P.D.L.R. and A.B. designed and performed research, analyzed the results and wrote the paper.

\section*{ACKNOWLEDGEMENTS}
The authors thank the Swiss
National Science Foundation for support under the grant 200021-138073 (S.A. and P.D.L.R.) and the 
Ambizione fellowship program (A.B.).

\end{document}